\documentclass[12pt]{article}
\normalbaselines
\catcode `\@=11
\@addtoreset{equation}{section}

\def\section{\@startsection {section}{1}{\z@}{-3.5ex plus -1ex minus
     -.2ex}{2.3ex plus .2ex}{\normalsize\bf}}
\def\subsection{\@startsection{subsection}{2}{\z@}{-3.25ex plus -1ex minus
 -.2ex}{1.5ex plus .2ex}{\normalsize\bf}}

\catcode `\@=12

\newcommand{\be}{\begin{equation}}
\newcommand{\ee}{\end{equation}}

\begin{document}

\title{\bf RELATIVISTIC SPINOR DYNAMICS INDUCING THE
EXTENDED LORENTZ-FORCE-LIKE EQUATION.}

\author{{\bf Andreas Bette}\\ 
\small{
The Royal Institute of Technology, KTH Syd, Campus Telge,
}
\\ \small{
Mariek{\"a}llgatan 3,
S-151 81 S{\"o}dert{\"a}lje,
Sweden.} \\ \small{e-mail: bette@kth.se
} 
\and {\bf {Jesus Buitrago}} \\ 
\small{Department of
Astrophysics of the University 
of La Laguna},  \\
\small{C/ Via Lactea, s/n, 
38205, La Laguna, 
Tenerife, Spain.} \\ \small{e-mail: jbg@ll.iac.es }
} \date{} \maketitle

\begin{abstract} 
\noindent
The special relativistic dynamical equation of the Lorentz force type
can be regarded as a consequence of a succession of space-time
dependent infinitesimal Lorentz transformations as shown by one of us
\cite{buitrago} and discussed in the introduction below. Such an
insight indicates that the Lorentz-force-like equation has an
extremely fundamental meaning in physics.  In this paper we therefore
present a set of dynamical Weyl spinor equations {\em inducing} the
extended Lorentz-force-like equation in the Minowski space-time.  The
term extended refers to the dynamics of some additional degrees of
freedom that may be associated with the classical spin namely with the
dynamics of three space-like mutually orthogonal four-vectors, all of
them orthogonal to the linear four-momentum of the object under
consideration.
\end{abstract}


\section{INTRODUCTION.}

Some years ago it has been noticed by J. Buitrago in \cite{buitrago}
that Lorentz force equation may be regarded as a consequence of the
action of infinitesimal Lorentz transformations on the velocity
four-vector of a relativistic particle, where the parameters of the
infinitesimal Lorentz transformations (i.e.\@ of infinitesimal boosts
and infinitesimal rotations) are regarded as functions of the position
of the particle and not just constants. If these infinitesimal
parameters are identified with the components of an external
electromagnetic field (evaluated at the four-position $x$ of the
particle) multiplied by the infinitesimal lapse of the particle's
proper time then the Lorentz force equation will follow automatically.

\vskip 10pt
\noindent
Let us reassume Buitrago's contribution. Consider the infinitesimal
Lorentz transformation in the Minkowski four-vector space $M_{v}$:

\begin{equation}\label{1}
u^{a}
(s+\Delta
s)=[\delta_{b}^{ \ a} + \Delta \ L^{a}_{b} (x(s))] u^{b}(s),
\end{equation}

\vskip 10pt
\noindent
where $u^{a}$ is a time-like or space-like Lorentz four-vector and
where $x(s)$, in the Minkow{\-}ski space-time $M$, denotes a trajectory
labelled by a parameter $s$.

\vskip 10pt
\noindent
The infinitesimal Lorentz transformation in (\ref{1}) is defined by:

\begin{equation}\label{2}
\Delta \ L^{a}_{b} 
(x(s)):= [\frac12 
\alpha^{c d} (x^{k}(s)) M_{cd \ \ b}^{\ \ a}
]
\Delta s,
\end{equation}

\vskip 10pt
\noindent
where $a,b,c,d$ indices refer to the Lorentz four-vector and
four-tensor character of the introduced quatities, so that e.g.\@
$\delta_{b}^{ \ a} $ is the Lorentz identity operator (matrix)
etc.. The six generators of the Lorentz transformations $M_{cd \ \
b}^{\ \ a}$ are, in the Minkowski four-vector space, defined by the
following constant Lorentz tensor of fourth rank\footnote{as
well-known from quantum mechanics the same generators multiplied by a
purely imaginary number e.g.\@ $i:=\sqrt{-1}$ are identified with the
intrinsic spin one angular four-momentum operator.}:

\begin{equation}\label{Lgen}
M_{cd \ \ b}^{\ \ a}=-M_{dc \ \ b}^{\ \ a} :=\delta_{c}^{\ a}
\ \eta_{db} - \delta_{d}^{\ a} \ \eta_{cb},
\end{equation}

\vskip 10pt
\noindent
where $\eta_{kl}$ denotes the Minkowski metric.  $\alpha^{cd}(x^{a}
({s})) \Delta s =-\alpha^{dc}(x^{a} ({s})) \Delta s $ represent the
infinitesimal parameters of the infinitesimal Lorentz transformations
changing continuously along the trajectory $x(s)$, while $ \Delta s $
measures the lapse of the parameter $s$ along the trajectory.  Using
(\ref{Lgen}) it is easy to see that:

\begin{equation}\label{3}
\frac12 
\alpha^{c d} (x^{k}(s)) 
M_{cd \ \ b}^{\ \ a}
=\frac12 \alpha^{c d} (x^{k}(s)) 
(\delta_{c}^{\ a}
\ \eta_{db} - \delta_{d}^{\ a} \ \eta_{cb})= \alpha^{a}_{\ \ b} (x^{k}(s)).
\end{equation}

\vskip 10pt
\noindent
Therefore the equation in (\ref{1}) may be now written as:

\begin{equation}\label{5}
\frac{u^{a}(s+\Delta s)-u^{a}(s)} {\Delta s} =\alpha^{a}_{\ \ b}
(x^{k}(s)) u^{b},
\end{equation}

\vskip 10pt
\noindent
where the identity in (\ref{3}) has been used. Taking the limit $\Delta s
\rightarrow 0$ in (\ref{5}) gives:

\begin{equation}\label{lorentz-like}
\frac{du^{a}}{ds}
={\alpha}^{a}_{\ \ b} (x^{k}(s)) 
u^{b}.
\end{equation}

\vskip 10pt
\noindent
It is easy to see from (\ref{lorentz-like}) that the necessarily
non-zero Lorentz norm of the four-vector $u^{a}$ is
preserved\footnote{because of $\alpha ' \textrm{s}$ antisymmetric
Lorentz tensor character.}  along the trajectory $x(s)$.

\vskip 10pt
\noindent
Two additional assumptions about $u^{a}$ namely that it is a time-like
four-vector and that it represents a four-velocity of a physical
massive system with a well-defined four-position $x$ in $M$ amounts
to\footnote{in this paper the signature convention of the Lorentz
metric in the Minkowski space $M$ and the corresponding Minkowski
vector space $M_v$ is $+---$.} the following conditions:

\begin{equation}\label{norm}
u^{a}=\frac{dx^{a}}{ds}
, \ \ u^{a}u_{a}=1.
\end{equation}

\vskip 10pt
\noindent
The two equations in (\ref{lorentz-like})-(\ref{norm}) can be regarded as a
second order Lorentz-force-like differential equation:

\begin{equation}\label{lorentz-like1}
\frac{d^{2}x^{a}}{ds^{2}}
={\alpha}^{a}_{\ \ b} (x^{k}) 
\frac{dx^{b}}{ds}.
\end{equation}

\vskip 10pt
\noindent
whose solutions are the assumed trajectories $x(s)$ in $M$, while the parameter $s$ 
is then recognised as the proper time of the system with the four-velocity $u^{a}$.

\vskip 10pt
\noindent
If we, in addition,  assume that the function $\alpha$, defining the infinitesimal
parameters of the infinitesimal Lorentz transformations as given in
(\ref{1}) and  (\ref{2}), is proportional to the external electromagnetic
field present along the system's trajectory in the following
simplest possible way:

\begin{equation}\label{4}
\alpha^{cd}(x^{m}(s)):=\frac{e}{m} F^{c d} (x^{m}(s)),
\end{equation}

\vskip 10pt
\noindent
then the equation in (\ref{lorentz-like1}) is not only
Lorentz-force-like but becomes exactly the Lorentz force equation.

\vskip 10pt
\noindent
The classical dynamical principle leading to the equation in
(\ref{lorentz-like}) where $u^{a}$ is a time-like or a space-like
Lorentz four-vector following a trajectory in $M$ is very simple and fully
geometrical.  The key point for its validity is the identity proved in
(\ref{3}).

\vskip 10pt
\noindent
At the classical relativistic level that we discuss in this
paper there is in this dynamical principle nothing to tell us how to
choose a second rank antisymmetric Lorentz tensor valued function
$\alpha$ in (\ref{2}).  We made a choice in (\ref{4}) and that
together with the two assumptions about $u^{a}$ in (\ref{norm}) 
produced the Lorentz force equation. However, whatever choice and
assumptions we make, quite generally, we get (\ref{lorentz-like})
from purely geometrical considerations. 

\vskip 10pt
\noindent
In order to be able to follow the ideas presented in the sequel, the
reader must have some basic knowledge about Weyl spinors and their
relation to the Lorentz tensors at least to the extent as presented on
the first pages of e.g.\@ \cite{hermosillo,pr1,jmp,prmc,prrw,Stewart}.
Basic knowledge of the philosophy behind Penrose's Twistor Theory can
also be of value when trying to understand the ideas that led us to
the results obtained in this paper.


\section{EXTENDED LORENTZ-FORCE-LIKE EQUATION.}

Consider the four Lorentz invariant, geometrically induced equations
such as in (\ref{lorentz-like}):

\begin{equation}\label{Pdot}
{\dot P}^{a}:= \frac{dP^{a}}{ds}=\alpha^{ab}(x^{c}(s)) \ P_{b},
\end{equation}

\begin{equation}\label{Sdot}
{\dot S}^{a}:= \frac{dS^{a}}{ds}=\alpha^{ab}(x^{c}(s)) \ S_{b},
\end{equation}

\begin{equation}\label{Vdot}
{\dot V}^{a}:= \frac{dV^{a}}{ds}=\alpha^{ab}(x^{c}(s)) \ V_{b},
\end{equation}

\begin{equation}\label{Wdot}
{\dot W}^{a}:= \frac{dW^{a}}{ds}=\alpha^{ab}(x^{c}(s)) \ W_{b},
\end{equation}

\vskip 10pt
\noindent
where $P$ is time-like while $S$, $V$ and $W$ are three space-like
Lorentz four-vectors following a trajectory $x(s)$ in $M$ and fulfilling
the following conditions along the trajectory:

\begin{equation}\label{orthnorm1}
m^{2}:={P}^{a}{ P}_{a} = - { S}^{a}{ S}_{a}=
- {V}^{a}{ V}_{a}=
- { W}^{a}{ W}_{a}
\not = 0,
\end{equation}

\begin{equation}\label{orthnorm2}
{S}^{a}{ P}_{a}=0, \ \ 
{V}^{a}{P}_{a}=0, \ \
{W}^{a}{ P}_{a}=0, \ \
{ S}^{a}{ W}_{a}=0, \ \
{S}^{a}{ V}_{a}=0, \ \
{W}^{a}{V}_{a}=0. 
\end{equation}

\vskip 10pt
\noindent
Note that $m^{2}$, the square of the norm of the time-like four-vector
$P$, is a constant of motion.  In (\ref{Pdot})-(\ref{orthnorm2}) we
thus defined dynamics of an orthogonal tetrad of Lorentz four-vectors
$P$, $S$, $V$ and $W$, following an, as yet unspecified, trajectory in
$M$, along which it is infinitesimally Lorentz transformed by an
external $\alpha$ field.

\vskip 10pt
\noindent
Now we make an additional assumption about the time-like four-vector
$P$ identifying it with the linear four-momentum of an object
following the trajectory $x(s)$ in $M$. Therefore we
require additionally:

\begin{equation}\label{xdot}
{\dot x}^{a}
:=\frac{dx^{a}}{ds}
 = \frac{{ P}^{a}}{\sqrt{ P^{b} P_{b}}}=\frac{{ P}^{a}}{m},
\end{equation}

\vskip 10pt
\noindent
so that the parameter $s$ may again be recognised as the proper time
of the object moving along the trajectory $x(s)$. This fact follows
trivially from (\ref{xdot}) because it implies that ${\dot x}^{a}
{\dot x}_{a} = 1$.

\vskip 10pt
\noindent
The equations in (\ref{Pdot}) and in (\ref{xdot}) define together the
(usual) Lorentz-force like equation while the additional equations in
(\ref{Sdot})-(\ref{Wdot}) obeying the conditions displayed in
(\ref{orthnorm1})-(\ref{orthnorm2}) form the extension of the
Lorentz-force-like equation alluded to in the title of this section.
This extension defines new degrees of freedom of the object following
the trajectory $x(s)$ in $M$. These degrees of freedom may be
associated with intrinsic spin of the object. See a short discussion
concerning this issue below. 

\vskip 10pt
\noindent
We wish also to stress the fact that the dynamical equations in
(\ref{Pdot})-(\ref{Wdot}), describing dynamics of the four four-vectors $P$,
$S$, $V$ and  $W$, are a consequence of the geometrical considerations as
briefly discussed in the introduction \cite{buitrago}.

\vskip 10pt
\noindent
To represent the four-vectors $P$, $S$, $V$ and $W$ we will use a pair
of non-proportional spinors $\pi$ and $\eta$. By doing so we will {\em
automatically} fulfil the conditions in
(\ref{orthnorm1})-(\ref{orthnorm2}).  Let therefore the time-like
four-vector $P$ and the three space-like four-vectors $S$, $V$, $W$ be
defined spinorially \cite{prrw} in the standard way (abstract index
notation in the sense of Penrose \cite{pr1} is used when appropriate):

\begin{equation}\label{P}
P_{a}:=P_{AA^{\prime}}={\pi}_{A^{\prime}} (s){\bar \pi}_{A}(s) 
+
{\bar  \eta}_{A} (s)
{ \eta}_{A^{\prime}}(s),
\end{equation}

\begin{equation}\label{S}
S_{a}:=S_{AA^{\prime}}={\pi}_{A^{\prime}} (s){\bar \pi}_{A}(s) 
-
{\bar  \eta}_{A} (s)
{ \eta}_{A^{\prime}}(s),
\end{equation}

\begin{equation}\label{V}
V_{a}:=V_{AA^{\prime}}={\pi}_{A^{\prime}} (s){\bar \eta}_{A}(s) 
+
{\bar  \pi}_{A} (s)
{ \eta}_{A^{\prime}}(s),
\end{equation}

\begin{equation}\label{W}
W_{a}:=W_{AA^{\prime}}=i({\pi}_{A^{\prime}} (s){\bar \eta}_{A}(s) 
-
{\bar  \pi}_{A} (s)
{ \eta}_{A^{\prime}}(s)).
\end{equation}

\vskip 10pt
\noindent
Using the definitions in (\ref{P})-(\ref{W}), simple spinor algebra
calculations show that the conditions in
(\ref{orthnorm1})-(\ref{orthnorm2}) are automatically
fullfilled. Certain attempts, to formulate the traditional Lorentz
force equation by the use of spinor representation of the linear
four-momentum $P$ as in (\ref{P}), have been put forward previously \cite{pleban}.

\vskip 10pt
\noindent
Now we claim that besides this automatic fulfillment of
(\ref{orthnorm1})-(\ref{orthnorm2}), all the equations in
(\ref{Pdot})-(\ref{Wdot}) 
are induced by the following relatively
simple dynamical spinor equations:

\begin{equation}\label{spinordynamicsproper1}
\frac{d { {\pi}}^{A^{\prime}}}{ds} \equiv {\dot {\pi}}^{A^{\prime}} =
-c \ {{\pi}}^{A^{\prime}} - b \ {\eta}^{A^{\prime}}, 
\ \
\frac{d { {\eta}}^{A^{\prime}}}{ds} \equiv
{\dot
{\eta}}^{A^{\prime}} = a \ {{\pi}}^{A^{\prime}} + c \
{\eta}^{A^{\prime}},
\end{equation}

\vskip 10pt
\noindent
where the complex valued Lorentz scalar functions $a$,  $b$  and $c$ are given by:

\begin{equation}\label{abc1}
{a}=\frac{{\alpha}^{S^{\prime}T^{\prime}}(x)
{{\eta}}_{S^{\prime}}
{{\eta}}_{T^{\prime}} }{({ \pi}^{K^{\prime}}{\eta}_{K^{\prime}} )}, \ \
{b}=\frac{{\alpha}^{S^{\prime} T^{\prime}}(x){{\pi}}_{S^{\prime}}
{{\pi}}_{T^{\prime}} }
{({ \pi}^{K^{\prime}}{\eta}_{K^{\prime}} )},
\  \
{c}=-\frac{{\alpha}^{S^{\prime} T^{\prime}}(x){{\pi}}_{S^{\prime}}
{{\eta}}_{T^{\prime}} }{({ \pi}^{K^{\prime}}{\eta}_{K^{\prime}} )}.
\end{equation}

\vskip 10pt
\noindent
and where the symmetric second rank spinor
${\alpha}^{S^{\prime}T^{\prime}}(x)$, in the standard manner,
represents the given external field $\alpha$ and where $x$ represent events
on the object's trajectory in $M$. 

\vskip 10pt
\noindent
If this claim is true then this implies that the equation in
(\ref{spinordynamicsproper1}) extends the geometrical principle as
described by J. Buitrago in \cite{buitrago} to the space of the two
spinors $\pi$, $\eta$. We now sketch the main parts of the proof
showing that (\ref{spinordynamicsproper1}) induces the equations in
(\ref{Pdot})-(\ref{Wdot}):

\vskip 10pt
\noindent
Multiplying the first equation in (\ref{spinordynamicsproper1}) by
${\bar {\pi}}^{A}$ and the second by ${\bar {\eta}}^{A}$ gives:

\begin{equation}\label{pidotspin}
{\bar {\pi}}^{A} 
{\dot {\pi}}^{A^{\prime}}
= {\bar {\pi}}^{A}  
(-c \ {{\pi}}^{A^{\prime}} - b \ {\eta}^{A^{\prime}}), 
\end{equation}

\begin{equation}\label{etadotspin}
{\bar {\eta}}^{A} 
{\dot {\eta}}^{A^{\prime}} = {\bar {\eta}}^{A} 
(a \ {{\pi}}^{A^{\prime}} + c \
{\eta}^{A^{\prime}}).
\end{equation}

\vskip 10pt
\noindent
Taking the complex conjugates of (\ref{pidotspin}) and (\ref{etadotspin}) gives:

\begin{equation}\label{pidotspinbar}
{{\pi}}^{A^{\prime}} {\dot {\bar {\pi}}}^{A} 
= {{\pi}}^{A^{\prime}} 
(-{\bar c} \  {\bar {\pi}}^{A} 
\ 
- {\bar b} \ {\bar {\eta}}^{A} 
), 
\end{equation}

\begin{equation}\label{etadotspinbar}
{{\eta}}^{A^{\prime}} 
{\dot {\bar {\eta}}}^{A}
= 
{{\eta}}^{A^{\prime}} 
({\bar a} {\bar \pi}^{A}+ {c} \ {\bar {\eta}}^{A} ).
\end{equation}

\vskip 10pt
\noindent
Adding the four equation in (\ref{pidotspin}), (\ref{etadotspin}),
(\ref{pidotspinbar}), (\ref{etadotspinbar}) sidewise to each other
gives:

$$
{\bar \pi}^{A} 
{\dot {\pi}}^{A^{\prime}} 
+ 
{\bar \eta}^{A} 
{\dot {\eta}}^{A^{\prime}} + 
 \textrm{c.c.}=
$$

\begin{equation}\label{Pdotspinabcadded}
=-[b {{\bar \pi}}^{A} {\eta}^{A^{\prime}} +
{\bar b} {{\bar \eta}}^{A} {\pi}^{A^{\prime}}
+ (c+{\bar c}){ {\bar \pi}}^{A} {\pi}^{A^{\prime}}]
+
[a {{\bar \eta}}^{A} {\pi}^{A^{\prime}} +
{\bar a} {{\bar \pi}}^{A} {\eta}^{A^{\prime}}
+ (c+{\bar c}){ {\bar \eta}}^{A} {\eta}^{A^{\prime}}].
\end{equation}

\vskip 10pt
\noindent
On the other hand we note that using spinor representation
of the equation in (\ref{Pdot}) gives:

\begin{equation}\label{Pdotspin}
{\dot {\bar \pi}}^{A} 
{\pi}^{A^{\prime}}
+ {\bar \pi}^{A} {{\dot \pi}^{A^{\prime}}}+
{\dot {\bar \eta}}^{A} 
{\eta}^{A^{\prime}}
+ {\bar \eta}^{A} {{\dot \eta}^{A^{\prime}}}
= \alpha^{AA^{\prime}BB^{\prime}} (x)
({\bar \pi}_{B}
{\pi_{B^{\prime}}}
+
{\bar \eta}_{B}
{\eta_{B^{\prime}}}),
\end{equation}

\vskip 10pt
\noindent
where $\alpha^{AA^{\prime}BB^{\prime}} (x)$ is the spinor equivalent
of the antisymmetric Lorentz-tensor $\alpha^{ab}$ in
(\ref{Pdot}). Spinor manipulating (\ref{Pdotspin}) further we note
that the external four-force field $\alpha$ in (\ref{Pdotspin}) may, in
the standard way, be represented by\footnote{algebraically in
(\ref{alpha}) the force field is represented by either an
antisymmetric Lorentz tensor of second rank or an hermitian spinor of
fourth rank twice primed and twice unprimed or equivalently by a
symmetric spinor of second rank either unprimed or primed, all these
representations being physically equivalent. We use therefore the same
generic letter $\alpha$ for these quantities.}:

\begin{equation}\label{alpha}
\alpha^{ab}(x)
=-\alpha^{ba}(x)=
\alpha^{AA^{\prime}BB^{\prime}}(x)
= -\alpha^{BB^{\prime}AA^{\prime}}(x):=
{\alpha}^{A^{\prime} B^{\prime}}(x) 
\
\epsilon^{AB}+ \textrm{c.c.}.
\end{equation}

\vskip 10pt
\noindent
Now we decompose the spinor $\alpha^{A^{\prime}B^{\prime}}$ as follows:

\begin{equation}\label{alphaAprimBprim}
{\alpha}^{A^{\prime} B^{\prime}
}(x) 
=
\frac{a}
{{\pi}^{C^{\prime}} \eta_{C^{\prime}}}
{\pi}^{A^{\prime}} \pi^{B^{\prime}}
+\frac{c}
{{\pi}^{C^{\prime}} \eta_{C^{\prime}}}
{\pi}^{A^{\prime}} \eta^{B^{\prime}} 
+\frac{c}
{{\pi}^{C^{\prime}} \eta_{C^{\prime}}} 
{\eta}^{A^{\prime}} \pi^{B^{\prime}} +
\frac{b}
{{\pi}^{C^{\prime}} \eta_{C^{\prime}}} 
{\eta}^{A^{\prime}} \eta^{B^{\prime}},
\end{equation}

\vskip 10pt
\noindent
with $a$, $c$, and $b$ being defined as in (\ref{abc1}).
Using the decomposition in (\ref{alphaAprimBprim}) we may
rewrite (\ref{Pdotspin})
according to\footnote{$c.c.$ is short hand notation for complex
conjugation.}:

$$
{\bar \pi}^{A} 
{\dot {\pi}}^{A^{\prime}} 
+ 
{\bar \eta}^{A} 
{\dot {\eta}}^{A^{\prime}} + 
 \textrm{c.c.}=
$$

\begin{equation}\label{Pdotspinabc}
=-[b {{\bar \pi}}^{A} {\eta}^{A^{\prime}} + {\bar b} {{\bar \eta}}^{A}
{\pi}^{A^{\prime}} + (c+{\bar c}){ {\bar \pi}}^{A} {\pi}^{A^{\prime}}]
+ [a {{\bar \eta}}^{A} {\pi}^{A^{\prime}} + {\bar a} {{\bar \pi}}^{A}
{\eta}^{A^{\prime}} + (c+{\bar c}){ {\bar \eta}}^{A}
{\eta}^{A^{\prime}}],
\end{equation}

\vskip 10pt
\noindent
proving our assertion that (\ref{Pdot}) is induced by
(\ref{spinordynamicsproper1}). This follows simply from the fact that
(\ref{Pdotspinabc}) and (\ref{Pdotspinabcadded}) are identical.

\vskip 10pt
\noindent
By imitating the above steps, using the spinor representations of $S$,
$V$, and $W$, as defined in (\ref{S}), (\ref{V}), (\ref{W}) it may be
shown that also the equations in (\ref{Sdot}), (\ref{Vdot}),
(\ref{Wdot}) are all induced by the {\em same} dynamical spinor equations in
(\ref{spinordynamicsproper1}). We call therefore these equations the
``master equations''. This completes the proof that (\ref{Pdot}),
(\ref{Sdot}), (\ref{Vdot}), (\ref{Wdot}) are induced by
(\ref{spinordynamicsproper1}) while the conditions in
(\ref{orthnorm1})-(\ref{orthnorm2}) are automatically fulfilled due to
the definitions in (\ref{P})-(\ref{W}).

\vskip 10pt
\noindent
If we require in addition that (\ref{xdot}), which when written
spinorially, reads :

\begin{equation}\label{xdotP}
{\dot x}^{AA^{\prime}}:= \frac{ \ \ dx^{AA^{\prime}}} {ds} =\frac{
{\pi}^{A^{\prime}} {\bar \pi}^{A} + {\bar \eta}^{A} {\eta}^{A^{\prime}} }
{\sqrt{2 \ ({\bar \pi}^{B} {\bar \eta}_{B} ) \ ({\pi}^{B^{\prime}}
{\eta}_{B^{\prime}} )}}=(\frac{P^{AA^{\prime}}}{m}),
\end{equation}

\vskip 10pt
\noindent
is valid with $x$, as always, denoting points (events) along the
object's trajectory in space-time, then the parameter $s$ is once
again recognised as the object's proper time parameter while $P$
denotes its linear momentum four-vector.

\vskip 10pt
\noindent
The $S$, $V$ and $W$ in (\ref{S})-(\ref{W}) may be thought of as
defining the axis of inertia rigidly attached to an object with the
linear four-momentum $P$. If so, then any space-like four-vector,
formed as a linear combination of $S$, $V$ and $W$ (divided by $m$ by
obvious dimensional requirements) defines the, so called,
Pauli-Luba{\'n}ski spin four-vector. Now, quick glance at any equation
arising as any such linear combination of the equations in
(\ref{S})-(\ref{W}) and at the equation (11.164) in \cite{jackson}
reveals that the two equations are proportional to each other only if
the gyromagnetic ratio $g$ is equal to two. The ``master equations''
in (\ref{spinordynamicsproper1}) together with the requirement in
(\ref{xdotP}) may thus be regarded as a classical relativistic limit
of the equations of motion that describe dynamics of the
(classical limit of the) spinning electron ($g=2$) with any constant
real value of its spin i.e.\@ with any value of the norm of its
Pauli-Luba{\'n}ski spin four-vector.

\vskip 10pt
\noindent
Note that the function defined by the positive real valued 
Lorentz  scalar function:

\begin{equation}\label{m}
P^{a}P_{a}= 2 \ ({\bar \pi}^{A} {\bar \eta}_{A} ) \ ({
\pi}^{A^{\prime}}{\eta}_{A^{\prime}} )=m^{2},
\end{equation}

\vskip 10pt
\noindent
should then  be identified with the square of the rest mass of the object
(particle) while the function:

\begin{equation}\label{|m|}
m= \sqrt 2 
\ \vert {\bar \pi}^{A} {\bar \eta}_{A} \vert 
= \sqrt 2 \
\vert {
\pi}^{A^{\prime}}{\eta}_{A^{\prime}} \vert,
\end{equation}

\vskip 10pt
\noindent
defines its positive rest mass.

\vskip 10pt
\noindent
Note also that ``master equations'' in (\ref{spinordynamicsproper1})
imply that not only the function\footnote{and thereby the function
$m^{2}$ in (\ref{m}).} $m \not = 0$, in (\ref{|m|}), is a constant of
motion but that entire complex valued Lorentz scalar function:

\begin{equation}\label{f}
f:= 
{
\pi}^{A^{\prime}}{\eta}_{A^{\prime}}=f_{0} \not = 0,
\end{equation}

\vskip 10pt
\noindent
is also a constant of motion. This may be easily proved because by contracting the first
equation in (\ref{spinordynamicsproper1}) with the spinor $\eta$ and
the second with the spinor $\pi$:

\begin{equation}\label{massconst1}
{{\eta}}_{A^{\prime}} \ {\dot {\pi}}^{A^{\prime}} = - {c}  \ f,
\end{equation}

\begin{equation}\label{massconst2}
{{\pi}}^{A^{\prime}} \ {\dot {\eta}}_{A^{\prime}} =  { c}  \ f,
\end{equation}

\vskip 10pt
\noindent
and by adding (\ref{massconst1}) and (\ref{massconst2}) to
each other, we find that the nonvanishing complex valued function $f$
in (\ref{f}) fulfils:

\begin{equation}\label{f1}
{\dot f} = 0 \ \ \textrm{i.e.} \ \ f=f_{0}=\textrm{const.} 
\ \ \vert f_{0} \vert > 0.
\end{equation}

\vskip 10pt
\noindent
In the next section we show explicitly how the ``master equations''
(\ref{spinordynamicsproper1}) can be integrated in the case when the
``magnetic'' and ``electric'' fields are constant, equally valued and
perpendicular to each other and in the case of constant ``electric''
field or constant ``magnetic'' field or both of them constant and
being parallel to each other (in some laboratory frame).

\vskip 10pt
\noindent
The obtained trajectories will, off course, be the very well known
ones, see e.g.\@ \cite{landau}. However we get some additional
information about the motion of the remaining legs of the tetrad
attached to our dynamical object. This may be interpreted as
precession of the intrinsic spin vector (including the kinematic
Thomas precession \cite{jackson}) attached to the object\footnote{an
alternative classical limit of the dynamics (of a ``charged''
relativistic spinning and massive object) that starts from a twistor
phase space formulation and the second order formulation of the
minimally coupled Dirac equation has been introduced in
\cite{hermosillo,ab5}.}.

\vskip 10pt
\noindent
Coping with the non-constant external $\alpha$ field is much more
difficult and we present spinor equations for this general case in the
last section of this paper. 

\section{TWO EXAMPLES AND THEIR SOLUTIONS.}

Concrete constructions of solutions require a choice of the external
field and a choice of a suitable frame (corresponding to a laboratory
frame in the Minkowski space-time) in the spinor space $S$.  The basis
in such a fixed (inertial) spin-frame (see e.g.\@ \cite{prrw,
Stewart}) will be denoted by $(\iota, o)$:

\begin{equation}\label{spinframe}
\iota_{A}, \   o_{A}
\ \ \textrm{and} \ \
{\bar \iota}_{A^{\prime}}, \
{\bar o}_{A^{\prime}},
\end{equation}

\vskip 10pt
\noindent
and normalized by the requirement:

\begin{equation}\label{spinframenorm}
\iota^{A} o_{A}=1, \ {\bar \iota}^{A^{\prime}}
{\bar o}_{A^{\prime}}=1.
\end{equation}

\vskip 10pt
\noindent
The two dynamical spinors $\pi$, $\eta$ may now, with respect to the
chosen fixed spin frame, be expressed by means of their components:

\begin{equation}\label{pi}
\pi^{A^{\prime}}= u
{\bar \iota}^{A^{\prime}}
-
z
{\bar o}^{A^{\prime}}, 
\ \
\eta^{A^{\prime}}= v
{\bar \iota}^{A^{\prime}}
- w
{\bar o}^{A^{\prime}},
\end{equation}

\begin{equation}\label{pibar}
{\bar \pi}^{A}= {\bar u}
{\iota}^{A}
-
{\bar z}
{o}^{A}, 
\ \
{\bar \eta}^{A}= 
{\bar v}
{ \iota}^{A}
- 
{\bar w}
{o}^{A},
\end{equation}

\vskip 10pt
\noindent
where, in order to diminish the abundance of the indices, for the
components of the two spinors $\pi$ and $\eta$ we introduced the
notation:
 
\begin{equation}\label{notation1}
\pi^{A^{\prime}}{\bar o}_{A^{\prime}}
= u,
\ \
\pi^{A^{\prime}}
{\bar \iota}_{A^{\prime}}=
z,
\ \ 
\eta^{A^{\prime}}{\bar o}_{A^{\prime}}
= v, \ \
\eta^{A^{\prime}}
{\bar \iota}_{A^{\prime}}=
w.
\end{equation}

\begin{equation}\label{notation2}
{\bar \pi}^{A}{o}_{A}
= 
{\bar u}
, 
\ \ 
{\bar \pi}^{A}{\iota}_{A}
= {\bar z}
,
\ \
{\bar \eta}^{A}{o}_{A}
= {\bar v}
,
\ \
{\bar \eta}^{A}{\iota}_{A}
= {\bar w}
.
\end{equation}

\vskip 10pt
\noindent
Using these coordinates the ``master equations'' in (\ref{spinordynamicsproper1}) become:

\begin{equation}\label{master}
{\dot z}=-cz-bw, \ \
{\dot w}=az+cw, \ \
{\dot u}=-cu-bv, \ \
{\dot v}=au+cv,
\end{equation}

\vskip 10pt
\noindent
while the spinor product defining the function $f$ (see (\ref{f})-(\ref{f1})) 
which is a constant of motion reads:

\begin{equation}\label{f0}
f
=zv-uw=f_{0}=\textrm{constant} \not = 0.
\end{equation}

\vskip 10pt
\noindent
The external Lorentz-force-like field $\alpha$ is, with
respect to the chosen spin frame $(\iota,  o)$ given by (see e.g. \cite{Stewart} p. 91):

\begin{equation}\label{alfa}
\alpha^{A^{\prime}
B^{\prime}}(x)=
\alpha_{0}(x) 
{\bar \iota}^{A^{\prime}} {\bar \iota}^{B^{\prime}} 
-2 \alpha_{1}(x) 
{\bar \iota}^{(A^{\prime}} {\bar o}^{B^{\prime})}  
+
\alpha_{2} (x)
{\bar o}^{A^{\prime}} {\bar o}^{B^{\prime}} .
\end{equation}

\vskip 10pt
\noindent
where we have: 

$$
\alpha_{0}(x)=\frac{(E_{x}-B_{y})+i(E_{y}+B_{x})}{2}, \ \
$$

\begin{equation}\label{EB}
\alpha_{1}(x)=-\frac{E_{z}+iB_{z}}{2}, \ \ 
\alpha_{2} (x)=-\frac{(E_{x}+B_{y})-i(E_{y}-B_{x})}{2}
\end{equation}

\vskip 10pt
\noindent
with ${\vec E}(x)=(E_{x},\ E_{y}, \ E_{z})$ and ${\vec B}(x)=(B_{x},\
B_{y}, \ B_{z})$ representing the applied Lorentz-force-like
field\footnote{Note that, in this context, our ''electric'' ($\vec E$)
and ''magnetic'' ($\vec B$) fields have absorbed the factor
$\frac{e}{m}$. In that way the dimension of the field $\alpha$, in
natural units, becomes inverse of the time.}.  Therefore from
(\ref{alfa}) it follows that the functions in (\ref{abc1}) are, in the
chosen frame, given by:

\begin{equation}\label{a}
{a}=\frac{\alpha_{0}w^{2}
-2\alpha_{1}
v w
+\alpha_{2} v^{2}
}{{f}_{0}},
\end{equation}

\begin{equation}\label{b}
{b}=\frac{\alpha_{0}z^{2}
-2\alpha_{1}
u z
+\alpha_{2}u^{2}
}{{f}_{0}},
\end{equation}

\begin{equation}\label{c}
{ c}= - \frac{\alpha_{0} z w
-\alpha_{1} u w
-\alpha_{1} v z
+\alpha_{2}u v
}{{f}_{0}}.
\end{equation}

\vskip 10pt
\noindent
where ${f}_{0}$ is the constant of motion obtained in (\ref{f1}).

\vskip 10pt
\noindent
Taking into the account the relation in (\ref{xdotP}) the components
of the linear momentum four-vector $P$ are, with respect to the chosen
spinor frame (that in the standard way defines the constant tetrad of
an inertial frame in the Minkowski four-vector space) given by:

\begin{equation}\label{E+Pz}
{P}^{AA^{\prime}}\iota_{A}{\bar\iota}_{A^{\prime}}
=\frac{m({\frac{dt}{ds}+\frac{dz}{ds}})}
{\sqrt{2}}
= 
\frac{E+p_{z}}{\sqrt{2}}=z{\bar z}+w{\bar w}, 
\end{equation}

\begin{equation}\label{E-Pz}
{P}^{AA^{\prime}}o_{A}{\bar o}_{A^{\prime}}
=\frac{m({\frac{dt}{ds}-\frac{dz}{ds}})}
{\sqrt{2}}=
\frac{E-p_{z}}{\sqrt{2}} =
u{\bar u}+v{\bar v}, 
\end{equation}

\begin{equation}\label{px+ipy}
{P}^{AA^{\prime}}o_{A}{\bar \iota}_{A^{\prime}}=
\frac{m({\frac{dx}{ds}+i\frac{dy}{ds}})}
{\sqrt{2}}=
\frac{p_{x}+ip_{y}}{\sqrt{2}}=
z{\bar u}+w{\bar v}, 
\end{equation}

\vskip 10pt
\noindent
where $m=\sqrt{2 f_{0} {\bar f}_{0}}$ and where $s$ is the proper time
of the system with the linear four-momentum $P$.

\vskip 10pt
\noindent
The three space-like four vectors $S$, $V$ and $W$ in
(\ref{S})-(\ref{W}) are, with respect to the chosen spinor frame,
given by:

\begin{equation}\label{S0Sz}
{S}^{AA^{\prime}}\iota_{A}{\bar\iota}_{A^{\prime}}
= 
\frac{S_{0}+S_{z}}{\sqrt{2}}
=z{\bar z}-w{\bar w}, \ \
{S}^{AA^{\prime}}o_{A}{\bar o}_{A^{\prime}}
=
\frac{S_{0}-S_{z}}{\sqrt{2}} =
u{\bar u}-v{\bar v}, 
\end{equation}

\begin{equation}\label{Sx+iSy}
{S}^{AA^{\prime}}o_{A}{\bar \iota}_{A^{\prime}}=
\frac{S_{x}+iS_{y}}{\sqrt{2}}=
z{\bar u}-w{\bar v}, 
\end{equation}

\begin{equation}\label{V0Vz}
{V}^{AA^{\prime}}\iota_{A}{\bar\iota}_{A^{\prime}}
= 
\frac{V_{0}+V_{z}}{\sqrt{2}}=z{\bar w}+w{\bar z}, 
\ \
{V}^{AA^{\prime}}o_{A}{\bar o}_{A^{\prime}}
=
\frac{V_{0}-V_{z}}{\sqrt{2}} =
u{\bar v}+v{\bar u}, 
\end{equation}

\begin{equation}\label{Vx+iVy}
{V}^{AA^{\prime}}o_{A}{\bar \iota}_{A^{\prime}}=
\frac{V_{x}+iV_{y}}{\sqrt{2}}=
z{\bar v}+w{\bar u}, 
\end{equation}

\begin{equation}\label{W_{0}+Wz}
{W}^{AA^{\prime}}\iota_{A}{\bar\iota}_{A^{\prime}}
=
\frac{W_{0}+W_{z}}{\sqrt{2}}=i(z{\bar w}-w{\bar z}), 
\ \
{W}^{AA^{\prime}}o_{A}{\bar o}_{A^{\prime}}
=
\frac{W_{0}-W_{z}}{\sqrt{2}} =
i(u{\bar v}-v{\bar u}), 
\end{equation}

\begin{equation}\label{Wx+Wy}
{W}^{AA^{\prime}}o_{A}{\bar \iota}_{A^{\prime}}=
\frac{W_{x}+iW_{y}}{\sqrt{2}}=
i(z{\bar v}-w{\bar u}). 
\end{equation}

\vskip 10pt
\noindent
If the external field $\alpha$ is constant i.e.\@ does not depend on
the position $x$ in $M$ then we may separate two cases: the first case
when the ``electric'' and ``magnetic'' parts of the $\alpha$ field are
equal in magnitude and perpendicular to each other (compare with the
solution of the problem 2 on page 58 in \cite{landau}) and the second
case when the only non-vanishing part of the $\alpha$ field is the
``electric'' part or only the ``magnetic'' part or both ``electric''
and the ``magnetic part are non-vanishing and parallel to each other
(the mathematics of these three, just mentioned, options is the
same so we call it case number two).

\vskip 10pt
\noindent
We now proceed to consider the case number one. For that reason we
note that (\ref{master}) and (\ref{a})-(\ref{c}) simplify if we
introduce the following constants and variables:

\begin{equation}\label{beta}
\beta_{0}
:=\frac{\alpha_{0}}{f_{0}}, \ \
\beta_{1}:=\frac{\alpha_{1}}{\alpha_{0}}, \ \ 
\beta_{2}:=\beta_{0}(\frac{\alpha_{2}}{\alpha_{0}}-(\frac{\alpha_{1}
}{\alpha_{0}
})^{2}),
\end{equation}

\begin{equation}\label{z1w1}
w_{1}:=w-\beta_{1} v, \ \
z_{1}:= z -\beta_{1} u, \ \
\end{equation}

\vskip 10pt
\noindent
The expressions in (\ref{a})-(\ref{c}) and in (\ref{f0}) now read:

\begin{equation}\label{a1}
{a}=\beta_{0}(w_{1})^{2}+\beta_{2}v^{2}, \   \
{b}=\beta_{0}(z_{1})^{2}+\beta_{2}u^{2},
\ \
{c}=-\beta_{0}z_{1}w_{1} - \beta_{2}uv,
\ \
{f_{0}}=z_{1}v - uw_{1}.
\end{equation}

\vskip 10pt
\noindent
We note further that:

\begin{equation}\label{beta20}
\beta_{2}=0   \ \
\textrm{implies that}  \ \
\vec E \cdot \vec B = {\vert \vec E \vert}^{2}-{\vert \vec B \vert}^{2} = 0.
\end{equation}

\vskip 10pt
\noindent
By putting $\beta_{2}=0 $ the expressions for the first three functions in (\ref{a1})
become considerably simplified and we obtain:

\begin{equation}\label{a2}
{a}=\beta_{0}(w_{1})^{2}, \ \
{b}=\beta_{0}z_{1}^{2}, \ \
{c}=-\beta_{0}z_{1}w_{1}.
\end{equation}

\vskip 10pt
\noindent
The dynamical equations in (\ref{master}) acquire now a form that is very
easy to solve:

\begin{equation}\label{dz1dw1}
\frac{d (z_{1}+\beta_{1}u)}{ds}=-\alpha_{0} \beta_{1} \  z_{1}, \ \
\frac{d (w_{1}+\beta_{1}v)}{ds}=-\alpha_{0} \beta_{1} \ w_{1},
\end{equation}

\begin{equation}\label{dudv}
\frac{d u}{ds}=-\alpha_{0}  \ z_{1}, \ \
\frac{d v}{ds}=-\alpha_{0}  \ w_{1}.
\end{equation}

\vskip 10pt
\noindent
Inserting the equations in (\ref{dudv}) into the equations in (\ref{dz1dw1}) yields:

\begin{equation}\label{dzdwbeta}
\frac{d z_{1}}{ds}  = 0, \ \ 
\frac{d w_{1}}{ds}  = 0. 
\end{equation}

\vskip 10pt
\noindent
Therefore for the constant external $\alpha$ field fulfilling the
condition in (\ref{beta20}) the following simple solutions for the
coordinates of the two spinors are obtained fom
(\ref{dudv})-(\ref{dzdwbeta}):

\begin{equation}\label{solution2a}
v=v_{0}-{\alpha_{0}} w_{10} \cdot s, \ \ 
u=u_{0}-{\alpha_{0}} z_{10} \cdot s,
\end{equation}

\begin{equation}\label{solution2b}
w=w_{10}+ \beta_{1} v_{0} - {\alpha_{1}} w_{10} \cdot s, \ \
z=z_{10}+ \beta_{1} u_{0} - {\alpha_{1}} z_{10} \cdot s.
\end{equation}

\vskip 10pt
\noindent
where the four complex numbers $z_{10}, \ w_{10}, \ u_{0}, \ v_{0}$ are
constants of the spinor motion. 

\vskip 10pt
\noindent
We choose now the $y$ axis along the
``electric'' field and the $z$ axis along the ``magnetic'' field and
denote their common value by $B=B_{z}=E_{y}$.  This gives that:

\begin{equation}\label{alphas}
 {\alpha_{1}}=-{\frac{iB}{2}}, \ \
 {\alpha_{0}}={\frac{iB}{2}}, \ \ 
 {\alpha_{2}}={\frac{iB}{2}}, \ \
 {\beta_{1}}=\frac{\alpha_{1}}{\alpha_{0}}=-1,
\end{equation}

\vskip 10pt
\noindent
while the solutions in (\ref{solution2a})-(\ref{solution2b}) may now be
written as follows:

\begin{equation}\label{solution2aa}
v=v_{0}-{\frac{iB}{2}}
w_{10} \cdot s, \ \ 
u=u_{0}-{\frac{iB}{2}}
z_{10} \cdot s,
\end{equation}

\begin{equation}\label{solution2bb}
w=w_{10}- v_{0}  + {\frac{iB}{2}} w_{10} \cdot
s, \ \ z=z_{10} - u_{0} + {\frac{iB}{2}} 
z_{10} \cdot s.
\end{equation}

\vskip 10pt
\noindent
Forming the components of the linear momentum $P$ in (\ref{P}) in the
chosen frame (using (\ref{E+Pz})-(\ref{px+ipy})) out of the spinor
components in (\ref{solution2aa})-(\ref{solution2bb}) gives:

\begin{equation}\label{E+pz}
\frac{E+p_{z}}{\sqrt{2}}=
(z_{10}-u_{0})({\bar z}_{10}- {\bar u}_{0})
+(w_{10}-v_{0})({\bar w}_{10}- {\bar v}_{0})
+
\end{equation}

$$
+\frac{iB}{2} 
({\bar z}_{10}u_{0} - {z}_{10}{\bar u}_{0}+
{\bar w}_{10}v_{0} - {w}_{10}{\bar v}_{0}) 
\cdot s + \frac{B^{2}}{4} ({\bar z}_{10}z_{10} + {w}_{10}{\bar w}_{10})
\cdot s^{2},
$$

\begin{equation}\label{E-pz}
\frac{E-p_{z}}{\sqrt{2}}=
(v_{0}{\bar v}_{0}+u_{0}{\bar u}_{0})
+\frac{iB}{2} 
({\bar z}_{10}u_{0} - {z}_{10}{\bar u}_{0}
{\bar w}_{10}v_{0} - {w}_{10}{\bar v}_{0}) 
\cdot s +
\frac{B^{2}}{4} ({\bar z}_{10}z_{10} + {w}_{10}{\bar w}_{10})
\cdot s^{2},
\end{equation}

\begin{equation}\label{Px+iPy}
\frac{p_{x}+ip_{y}}{\sqrt{2}}=
({\bar u}_{0}z_{10} -
{u}_{0}{\bar u}_{0}+
{\bar v}_{0}w_{10} - {v}_{0}{\bar v}_{0})
+\frac{iB}{2} 
({\bar z}_{10}z_{10} + {w}_{10}{\bar w}_{10}
-u_{0}{\bar z}_{10}+
\end{equation}

$$ 
+z_{10}{\bar u}_{0} - {\bar w}_{10}v_{0} + {w}_{10}{\bar v}_{0})
\cdot s - \frac{B^{2}}{4} ({\bar z}_{10}z_{10} + {w}_{10}{\bar w}_{10})
\cdot s^{2}.
$$

\vskip 10pt
\noindent
Substracting (\ref{E-pz}) 
from (\ref{E+pz}) gives that $p_{z}$ is a constant of motion:

$$ p_{z}=\frac{(z_{10}-u_{0})({\bar z}_{10}- {\bar
u}_{0})+(w_{10}-v_{0})({\bar w}_{10}- {\bar v}_{0})-(v_{0}{\bar
v}_{0}+u_{0}{\bar u}_{0})}{\sqrt{2}}=\textrm{const.}
$$

\vskip 10pt
\noindent
Adding (\ref{E-pz}) to (\ref{E+pz}) and to (\ref{Px+iPy}) 
and to its complex
conjugate reveals that $E+p_{x}$ is also a constant of motion:

$$ E+p_{x}=\frac{2 \Re ({\bar u}_{0}z_{10}+{\bar v}_{0}w_{10})-
({\bar u}_{0}u_{0}+{\bar v}_{0}v_{0})+(z_{10}-u_{0})({\bar z}_{10}-{\bar u}_{0})}{\sqrt{2}}=\textrm{const.}
$$

\vskip 10pt
\noindent
$\Re$ is short notation for the real part of.  The trajectory in
space-time is now obtained by a simple integration of
(\ref{E+pz})-(\ref{Px+iPy}) using the relation between the position
four-vector and the linear four-momentum as displayed in
(\ref{E+Pz})-(\ref{px+ipy}). The trajectories coincide qualitatively
with the ones obtained in \cite{landau} as they off course should.

\vskip 10pt
\noindent
Inserting the solutions in (\ref{solution2aa})-(\ref{solution2bb})
into the definitions in (\ref{S})-(\ref{W}) describes the motion
(precession) of the tetrad attached to the object. It solves thereby
the dynamical equation in (\ref{Sdot})-(\ref{Wdot}) subject to the
condition in (\ref{orthnorm1})-(\ref{orthnorm2}). The precession
consists of two parts one purely kinematical (the Thomas precession)
and one dynamical induced by the ``master equation''. 

\vskip 10pt
\noindent
We conclude that the first special case has a simple explicit
analytical solution.

\vskip 10pt
\noindent
Now we proceed to analyse the second case when the ``magnetic'' part
and the ``electric'' part of the $\alpha$ field are parallell to each
other or any of them is vanishing. At the end we specilize to the case
of constant ``magnetic'' field that gives a very well-known circular
trajectory, however, the ``master equations'' imply also the
precession of the legs of the tetrad in (\ref{S})-(\ref{W}). Now we
proceed to the construction of the explicit solution. Choose therefore
the $z$ axis as the common line along which the (possibly) two fields
(``magnetic'' and ``electric'') are directed.  Then the formulas for
the functions in (\ref{a})-(\ref{b}) simplify again and read:

\begin{equation}\label{abc3}
{a}=-\frac{
2\alpha_{1}
v w
}{{f}_{0}}, \ \
{b}=-\frac{
2\alpha_{1}
u z
}{{f}_{0}},
\ \
{ c}=  \frac{
\alpha_{1} u w
+\alpha_{1} v z
}{{f}_{0}}.
\end{equation}

\vskip 10pt
\noindent
Inserting the functions in (\ref{abc3}) into the ``master equation''
in (\ref{master}) and  using (\ref{f0}) give a trivially
simple set of four first order equations for the four components
of the two spinors:

\begin{equation}\label{master1}
{\dot u}=\alpha_{1} u, \ \
{\dot z}=-\alpha_{1} z, \ \
{\dot v}=\alpha_{1} v, \ \
{\dot w}=-\alpha_{1} w.
\end{equation}

\vskip 10pt
\noindent
The solutions of (\ref{master1}) are of course given by:

\begin{equation}\label{solution1}
{u}=u_{0}e^{\alpha_{1} s}, \ \ {z}=
z_{0}
e^{-\alpha_{1}s},
\ \
{v}=v_{0}e^{\alpha_{1}s}, \ \
{w}=w_{0} e^{-\alpha_{1}s},
\end{equation}

\vskip 10pt
\noindent
where $u_{0}$, $z_{0}$, $v_{0}$ and $w_{0}$ are complex valued
constants.  Inserting (\ref{solution1}) into the expressions in
(\ref{E+Pz})-(\ref{Wx+Wy}) gives final solution for this case.
Therefore we may conclude that the second case with the constant
external ``magnetic'' and ``electric'' field is also explicitly
solved.

\vskip 10pt
\noindent
Exploring the second solution a bit further we note that in the special
case of pure constant ``magnetic'' field along the $z$ axis, since $
\alpha_{1}=-iB_{z}/2$, the integration constants $u_{0}$, $z_{0}$, $v_{0}$
and $w_{0}$ may be chosen as real.  From (\ref{Sx+iSy}) we find that $ S_x $
and $ S_y $ precess around the $ z $ axis according to

\begin{equation}
S_x= \sqrt{2} (z_0 u_0 - w_0 v_0) \cos B_{z} s,
\end{equation}

\begin{equation}
S_y= \sqrt{2} (z_0 u_0 - w_0 v_0) \sin B_{z} s,
\end{equation}

\vskip 10pt
\noindent
with similar behaviour for $ V_x, V_y, W_x $ and $ W_y $. On the other
hand, the $ S_z $ , $V_z$ and $W_z$ components remain  constant,
independent of the strength of the magnetic field, given by

\begin{equation}
S_z = \frac{\sqrt{2}}{2} (z_0^2+v_0^2-w_0^2-u_0^2)
\end{equation}

\begin{equation}
V_z = {\sqrt{2}}(z_0 v_0 - u_0 v_0), \ \ W_{z} = 0.
\end{equation}


\section{SPINOR DYNAMICS.}

If the external field $\alpha$ is not constant in space-time, the
integration of the ``master equations'' in
(\ref{spinordynamicsproper1}) cannot be performed directly, as done in
the previous section, because the functions $a$, $b$ and $c$ depend
explicitly on the object's position four-vector $x$.  To remedy this,
we let the position four-vector of the object be also translated into
its spinorial form by the following construction\footnote{this is
simply a projection of the position four-vector on the four legs of
the dynamical tetrad defined by the four four-vectors $P$, $S$, $V$
and $W$.}:

\begin{equation}\label{x}
x^{a}:=x^{AA^{\prime}}= h_{1}(s) 
\
{\pi}^{A^{\prime}} {\bar \pi}^{A} + h_{2}(s)  \ \
{\bar \eta}^{A} { \eta}^{A^{\prime}} + h_{3}(s) \
{\pi}^{A^{\prime}}{\bar \eta}^{A} + {\bar h}_{3}(s) \ {\bar \pi}^{A}
{{\eta}}^{A^{\prime}},
\end{equation}

\vskip 10pt
\noindent
where $h_{1}$ and $h_{2}$ are real valued Lorentz, $s$ parameter
dependent, scalar functions and where $h_{3}$ is a complex valued
Lorentz, $s$ parameter dependent, scalar function. 

\vskip 10pt
\noindent
Contracting\footnote{note that assuming the validity of (\ref{xdotP})
amounts to identification of the parameter $s$ with the proper time of
the object.} (\ref{xdotP}) with the two spinors ${\bar \eta}$
and ${\bar \pi}$ gives:

\begin{equation}\label{h1}
{\bar \eta}_{A}{\dot x}^{AA^{\prime}}=\frac{{\bar f} {\pi}^{A^{\prime}}} 
{\sqrt{2 \ ({\bar \pi}^{B} {\bar \eta}_{B})( {\pi}^{B^{\prime}} {\eta}_{B^{\prime}})}},
\end{equation}

\begin{equation}\label{h2}
{\bar \pi}_{A}{\dot x}^{AA^{\prime}}=-\frac{{\bar f} {\eta}^{A^{\prime}}} 
{\sqrt{2 \ ({\bar \pi}^{B} {\bar \eta}_{B})( {\pi}^{B^{\prime}} {\eta}_{B^{\prime}})}}.
\end{equation}

\vskip 10pt
\noindent
Taking derivative of (\ref{x}) with respect to $s$ gives:

\begin{equation}\label{xderivative}
{\dot x}^{AA^{\prime}}= {\dot h}_{1}(s) 
\
{\pi}^{A^{\prime}} {\bar \pi}^{A} + {\dot h}_{2}(s)   
\ \
{\bar \eta}^{A} { \eta}^{A^{\prime}} + {\dot h}_{3}(s) 
\
{\pi}^{A^{\prime}}{\bar \eta}^{A} + {\dot {\bar h}}_{3}(s) \ {\bar \pi}^{A}
{{\eta}}^{A^{\prime}}+ 
\end{equation}

$$
+ h_{1}(s) 
\
{{\dot \pi}}^{A^{\prime}} {\bar \pi}^{A} + h_{2}(s)  \ \
{\bar \eta}^{A} {{\dot \eta}}^{A^{\prime}} + h_{3}(s) \
{{\dot \pi}}^{A^{\prime}}{\bar \eta}^{A} + {\bar h}_{3}(s) \ {\bar \pi}^{A}
{{{\dot \eta}}}^{A^{\prime}}+
$$

$$
+h_{1}(s) 
\
{\pi}^{A^{\prime}} {\dot {\bar \pi}}^{A} + h_{2}(s)  \ \
{\dot {\bar \eta}}^{A} { \eta}^{A^{\prime}} + h_{3}(s) \
{\pi}^{A^{\prime}}{\dot {\bar \eta}}^{A} + {\bar h}_{3}(s) \ {\dot {\bar \pi}}^{A}
{{\eta}}^{A^{\prime}}.
$$

\vskip 10pt
\noindent
Inserting the right hand sides of the ``master equtions''
(\ref{spinordynamicsproper1}) (and their complex conjugates) into
(\ref{xderivative}), contracting\footnote{it is easier first to
contract and later to insert the ``master equations''.} the obtained
result at first with ${\bar \eta}_{A}$ and thereafter with ${\bar
\pi}_{A}$ yield two expressions that can be compared with right hand side
of (\ref{h1})-(\ref{h2}). By that we obtain the following four first order
differential equations for the scalar functions $h_{1}(s)$,
$h_{2}(s)$, $h_{3}(s)$:

$$
\frac{dh_{1}}{ds}-h_{1}(c+{\bar c})+h_{3}{\bar a}+{\bar h}_{3}{a}=\frac{1}{m},
\ \ \
\frac{dh_{2}}{ds}+h_{2}(c+{\bar c})-(h_{3}{b}+{\bar h}_{3}{\bar b})=\frac{1}{m},
$$

\begin{equation}\label{dot}
\frac{dh_{3}}{ds}-h_{1}{\bar b} + h_{2}{a} - h_{3}{c} + {h}_{3}{\bar c}=0.
\end{equation}

\vskip 10pt
\noindent
``Master equations'' in (\ref{spinordynamicsproper1}) and the
equations in (\ref{dot}) imply that the parameter $s$ is the proper
time of the object. Together they form a closed system of first order
differential equations that induces the equations in
(\ref{Pdot})-(\ref{Wdot}) fulfilling
(\ref{orthnorm1})-(\ref{orthnorm2}) and the physical requirement
(\ref{xdot}).

\vskip 10pt
\noindent
Note that space-time dynamics described in this abstract way (entirely
by the equations in (\ref{spinordynamicsproper1}) and (\ref{dot})) do
not make use of the notion of the space-time manifold at all. Position
events traced out by a relativistic system become secondary
constructions.

\section{SUMMARY AND REMARKS.}

According to the principle presented in \cite{buitrago} and summarised
in the introduction, any Lorentz force-like equation can be regarded
as a consequence of the geometry of the Minkowski four-vector
space. This principle is here extended to the spinor space and allows
us to discover a set of coupled spinor equations that describe
dynamics of a massive and spinning classical object (with g=2). The
solutions of these equations describe not only the world trajectories
of the object under study but also the degrees of freedom that can be
associated to its intrinsic classical (limit of its quantum mechanical
discrete) continuous spin values. It would be very interesting to know
whether it is possible to find a Lorentz invariant
hamiltonian/lagrangian formulation from which the equations in
(\ref{spinordynamicsproper1}) and in (\ref{dot}) can be derived. If it
is possible then a subsequent quantisation procedure would produce a
Lorentz invariant first quantised relativistic theory describing a
massive, spinning (for any quantised value of the spin) object acted
upon by an applied external $\alpha$. These speculations, however, we
postpone to future investigations.

\section*{ACKNOWLEDGMENTS.}

One of the authors (A.B.) would like to thank ``IAC'' (Instituto de
Astrofisica de Canarias in Tenerife, Spain) for the financial
support. He also wishes to thank The Royal Institute of Technology-KTH
Syd for providing him with the so called ``fakir'' funds that freed
him from the teaching duties at the time of this research.  We also
thank Professors Evencio Mediavilla and the members of his group for
their hospitality at IAC, where this work was partly carried out and
finally accomplished within the frame of the IAC project number 6/88.

\end{document}